\documentclass[aps,prl,reprint,groupedaddress]{revtex4-1}

\usepackage{amsmath}
\usepackage{graphicx,psfrag}
\usepackage[caption=false]{subfig}
\usepackage{verbatim}
\begin{document}
  

\title{Large fluctuations in non-ideal coarse-grained systems}


\author{M. Reza Parsa}
\email[]{mparsa@ucmerced.edu}
\affiliation{Department of Applied Mathematics, University of California, Merced, CA 95343, USA}

\author{Alexander J. Wagner}
\email[]{alexander.wagner@ndsu.edu}
\homepage[]{www.ndsu.edu/pubweb/$\sim$carswagn}
\affiliation{Department of Physics, North Dakota State University, Fargo, North Dakota 58108, USA}


\date{\today}

\begin{abstract}
 Using the recently introduced Molecular Dynamics Lattice Gas (MDLG) approach, we test fluctuations of coarse-grained quantities. We show that as soon as the system can no longer be considered an ideal gas fluctuations fail to diminish upon coarse-graining as is usually expected. These results suggest that current approaches to simulating fluctuating hydrodynamics may have to be augmented to achieve quantitative results for systems with a non-ideal equation of state. The MDLG method gives a guidance to the exact nature of the fluctuation in such systems. 
\end{abstract}


\maketitle

Predicting fluctuations at small scales for coarse-grained systems can be challenging\cite{hoyt2001method,kocher2016incorporating,altaner2012fluctuation}. Typically theoretical predictions exist for fluctuations in the hydrodynamics (i.e. long wavelength) limit\cite{grosfils1992spontaneous} that can be used to tune fluctuation terms, which is the basis of the Langevin approach to fluctuations \cite{1997AmJPh..65.1079L}. Fluctuating terms in the governing equations are, however, usually assumed to be spatially uncorrelated, so that the hydrodynamic behavior has to emerge from the local fluctuations that are introduced into the system. To complicate the situation fluctuations are often most relevant at scales much smaller than the hydrodynamic length-scales. This makes it imperative to develop methods that are able to reproduce correct fluctuations at much smaller length-scales\cite{kocher2016incorporating}. It is, however, generally not known what form the small-scale fluctuations for such a coarse-grained system should take. This is a general problem that arises in most simulation contexts where fluctuations are important.

One obvious way to address this shortfall is to use information from a microscopic model, e.g. a Molecular Dynamics (MD) simulation, to measure the small scale fluctuations\cite{hoyt2001method}. This information can then be used to tune the fluctuating terms in a coarse-grained simulation method \cite{kocher2016incorporating}. In this letter we show that it is possible to go a step further and use a coarse-graining mapping between a MD simulation and a coarse-grained mesoscopic model to directly measure the correct fluctuations of the model variables for different coarse-graining scales. Our results indicate that fluctuations of the model variables can be orders of magnitude larger than was previously expected. 

In this letter we focus on predictions for fluctuating lattice Boltzmann approaches, although we would like to stress that the mapping approach is far more general and could be applied to most other coarse-grained models. For the special case of an ideal gas fluctuations are fairly well understood. This is why this is usually taken as a starting point, e.g for fluctuating lattice Boltzmann \cite{ladd1993short,adhikari2005fluctuating,dunweg2007statistical,kaehler2013fluctuating}. For other systems fluctuations arise from the discrete nature of the representation as in lattice gases \cite{grosfils1992spontaneous, blommel2018integer} or Stochastic Rotation Dynamics \cite{ihle2003stochastic,tuzel2006dynamic}. Other discrete versions like Dissipative Particle Dynamics \cite{hoogerbrugge1992simulating,espanol1995hydrodynamics,vazquez2009consistent} includes tunable fluctuating forces. For non-ideal systems, however, it is typically less clear what the correct fluctuations should look like \cite{gross2010thermal,ollila2011fluctuating,thampi2011lattice,belardinelli2015fluctuating}. Because of this difficulty we developed a direct mapping from Molecular Dynamic (MD) onto a lattice gas (MDLG) \cite{parsa2017lattice} where fluctuations in a non-ideal coarse-grained system can be easily observed. In this letter we show the results of applying MDLG to analyzing equilibrium fluctuations at different densities of a system of Lenard-Jones particles.  

The direct mapping between MD and a lattice gas, leads to an integer lattice gas. Integer lattice gases exist, but are somewhat rare \cite{Molvig88,chen1997digital,teixeira1997digital,boghosian1997integer,chopard1998multiparticle,blommel2018integer}. Blommel showed that an integer lattice gas can closely model the fluctuations of an ideal gas \cite{blommel2018integer}. Such lattice gases have occupation numbers $n_i(x,t)$ as their fundamental variables. They are defined on lattice points $x$ and are associated with lattice velocities $v_i$ such that $x+v_i$ is again a lattice position. The evolution equation of a lattice gas can be written as
\begin{equation}
  n_i(x+v_i,t+\Delta t)=n_i(x,t)+\Xi_i
\end{equation}
where the $\Xi_i$ is the lattice gas collision operator. We also define the number density $\rho=\sum_i n_i$. For the integer lattice gas of Blommel \cite{blommel2018integer} the $n_i$ follow the expectation for an ideal gas\cite{landau1969statistical} and are Poisson distributed
\begin{equation}
  P(n_i)=\exp(-f_i^{eq})(f_i^{eq})^{n_i}/n_i!
\end{equation}
where $f_i^{eq}=\langle n_i\rangle$. The number density $\rho$ is similarly Poisson distributed with average $\rho^{eq}=\sum_i f_i^{eq}$. 

\begin{figure}
  \centering
  \subfloat[]{\includegraphics[width=0.9\columnwidth]{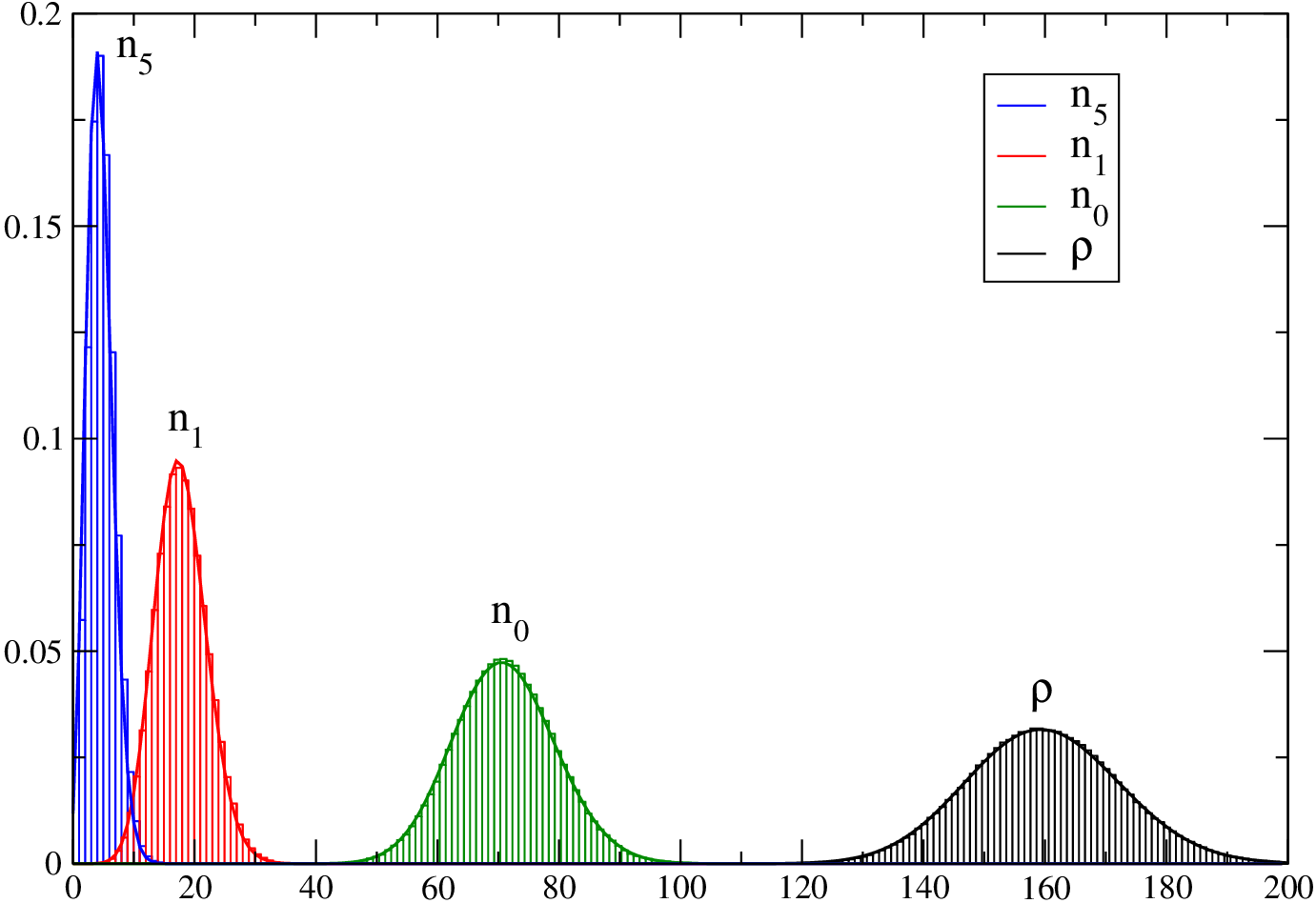}}\\
  \subfloat[]{\includegraphics[width=0.9\columnwidth]{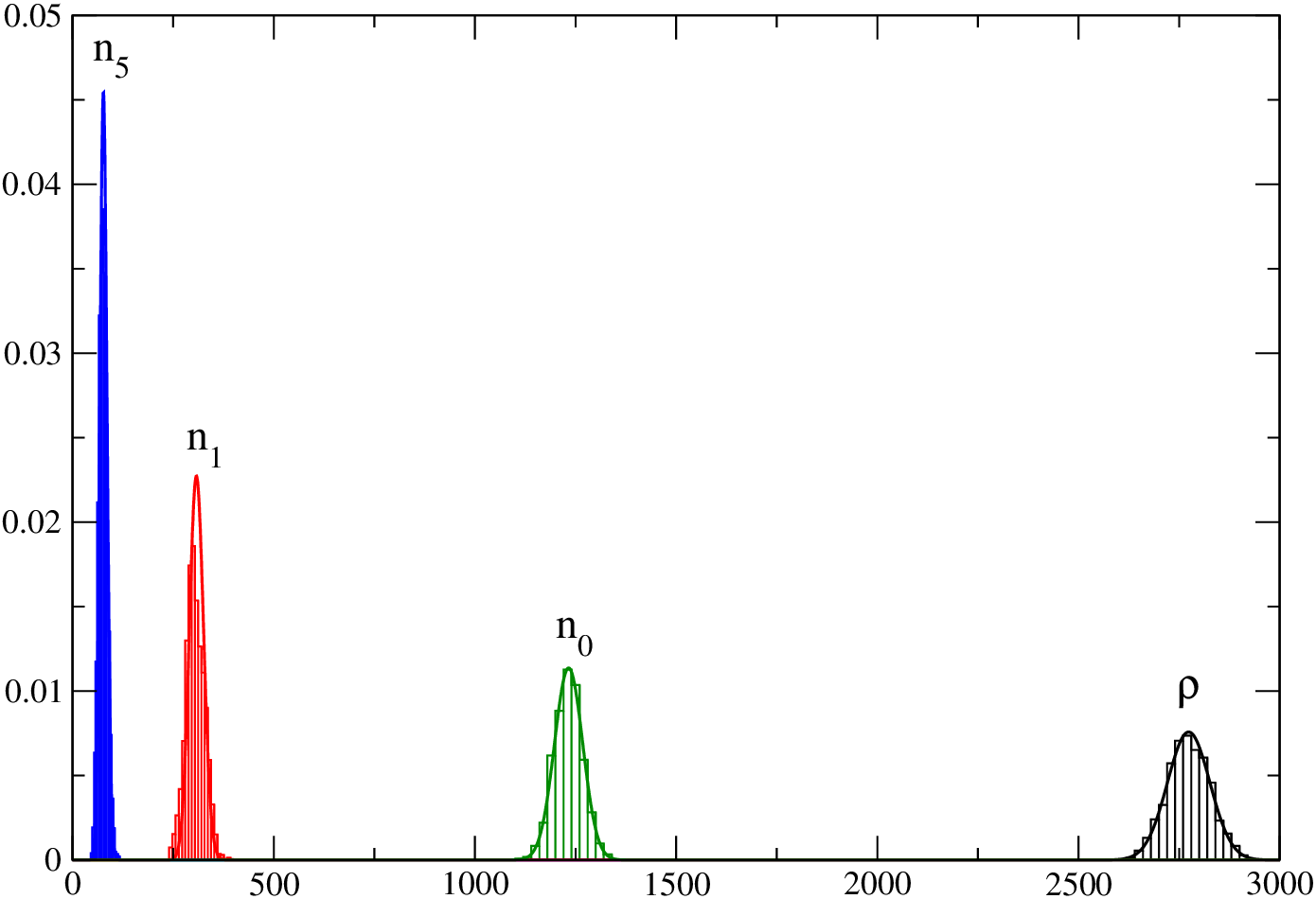}}
  \caption{Distribution of the MDLG occupation numbers $n_i$ and the total density (histograms) compared to the Poisson distribution (solid lines) for number density $\rho\approx 0.011$ for different lattice spacings (a) $\Delta x= 120$ and (b) $\Delta x= 500$.}
  \label{fig1}
\end{figure}

To examine the fluctuation behavior of a non-ideal system (as represented by a Molecular Dynamics simulation) we employ the MDLG approach. Here a lattice is superimposed on the Molecular Dynamics simulation and the displacement of particles from one lattice cell to another a distance $v_i$ away is during a timestep $\Delta t$ is then identified with the lattice gas occupation number $n_i$. This provides us with a lattice gas that exactly represents the underlying Molecular Dynamics system \cite{parsa2017lattice}.
Explicitly, if $x_j(t)$ is the position of the $j$th particle at time $t$, we can write
\begin{equation}
  n_i(x,t)=\sum_{j=1}^N \Delta_{x}[x_j(t)] \Delta_{x-v_i}[x_j(t-\Delta t)]
\end{equation}
where $\Delta_{y}[x]$ is one if $x_\alpha\in [y_\alpha,y_\alpha+\Delta x]$ and zero otherwise for all coordinates $\alpha$ and $N$ is the total number of particles.  

We have previously shown that such an MDLG can have an equilibrium distribution that is equivalent to the standard lattice Boltzmann equilibrium distributions when the combination of the time step $\Delta t$ and lattice size $\Delta x$ is chosen such that
\begin{equation}
  a^2 = \frac{\langle \delta(\Delta t)^2\rangle}{d\Delta x^2} \approx 0.18 \approx \frac{1}{6}
  \label{a2eqn}
\end{equation}
where $\delta(\Delta t)$ represents the displacement of a MD particle in the simulation for a time-interval of $\Delta t$ and $d$ is the number of spatial dimensions \cite{parsa2018validity}. For this particular choice the equilibrium distribution of the lattice gas recovers the standard D2Q9 equilibrium distribution of lattice Boltzmann for small velocities \cite{qian1992lattice}. While the MDLG approach will generate correct results for all values for $a^2$ the approach only recovers the standard lattice Boltzmann approach for $a^2\approx 1/6$. To ensure comparability of results to fluctuating lattice Boltzmann implementations \cite{ladd1993short,adhikari2005fluctuating,gross2010thermal,ollila2011fluctuating} we restrict ourselves to this value here. 

We now examine the equilibrium fluctuations of this lattice gas. We start with a dilute gas, for which we expect the assumptions of ideal gas fluctuations to hold to good approximation. We perform a Molecular Dynamics simulation with $99856$ particles on a $3000\times 3000$ lattice (length measured in LJ units) at a high temperature of 50, also in LJ units as in \cite{parsa2017lattice}, to prevent phase-separation. If we take the effective radius for excluded volume at about $r_c=0.75$ [corresponding to $1/2 k_BT=V(r_c)$], this translates into a volume fraction of $\phi=4.9\;10^{-3}$. For this volume fraction we indeed find good agreement between the distribution of the lattice gas occupation numbers and the Poisson distribution, as is shown in Figure \ref{fig1} for two different lattice spacings $\Delta x$.

\begin{figure}
  \centering
  \subfloat[]{\includegraphics[width=0.9\columnwidth]{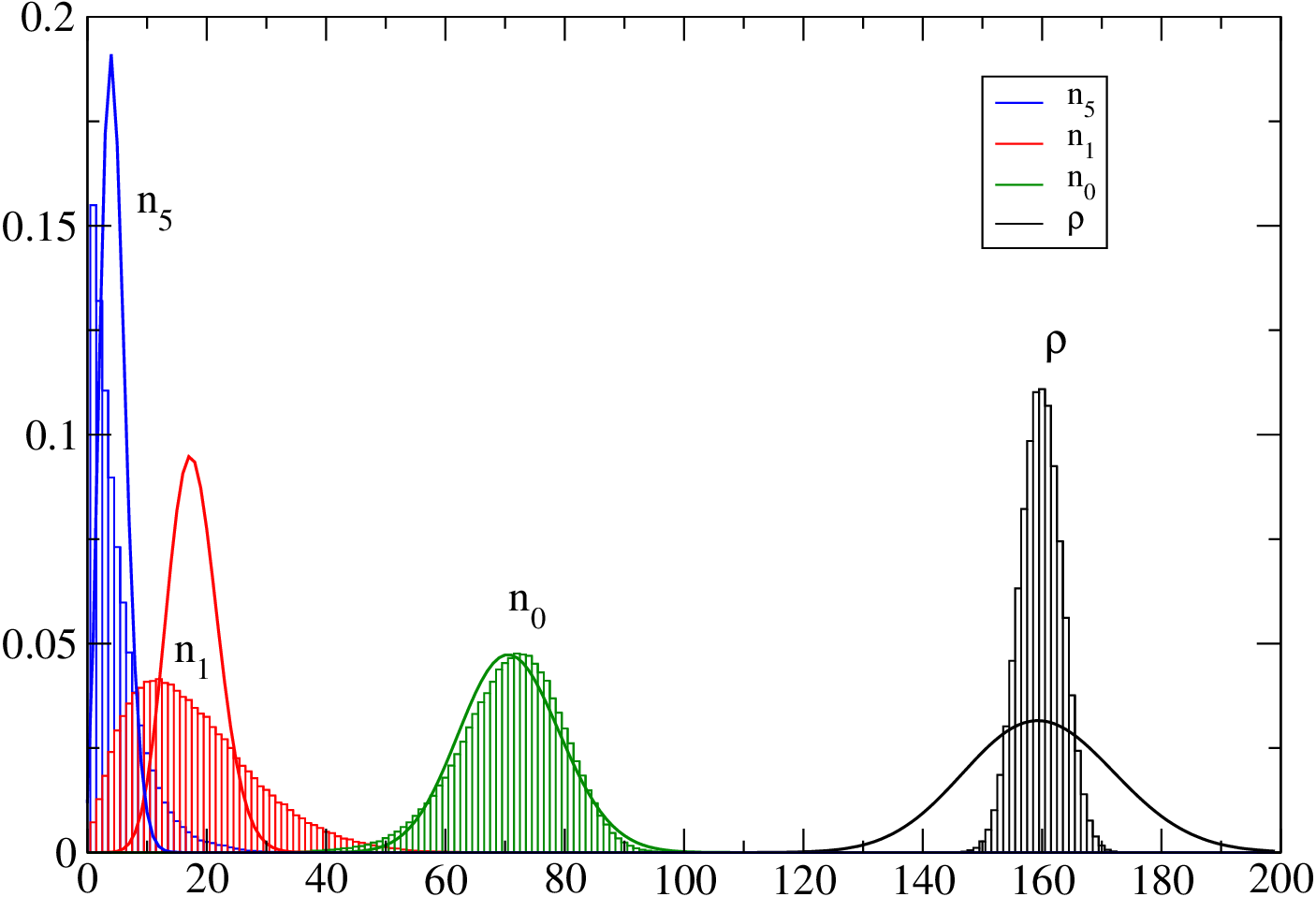}}\\
  \subfloat[]{\includegraphics[width=0.9\columnwidth]{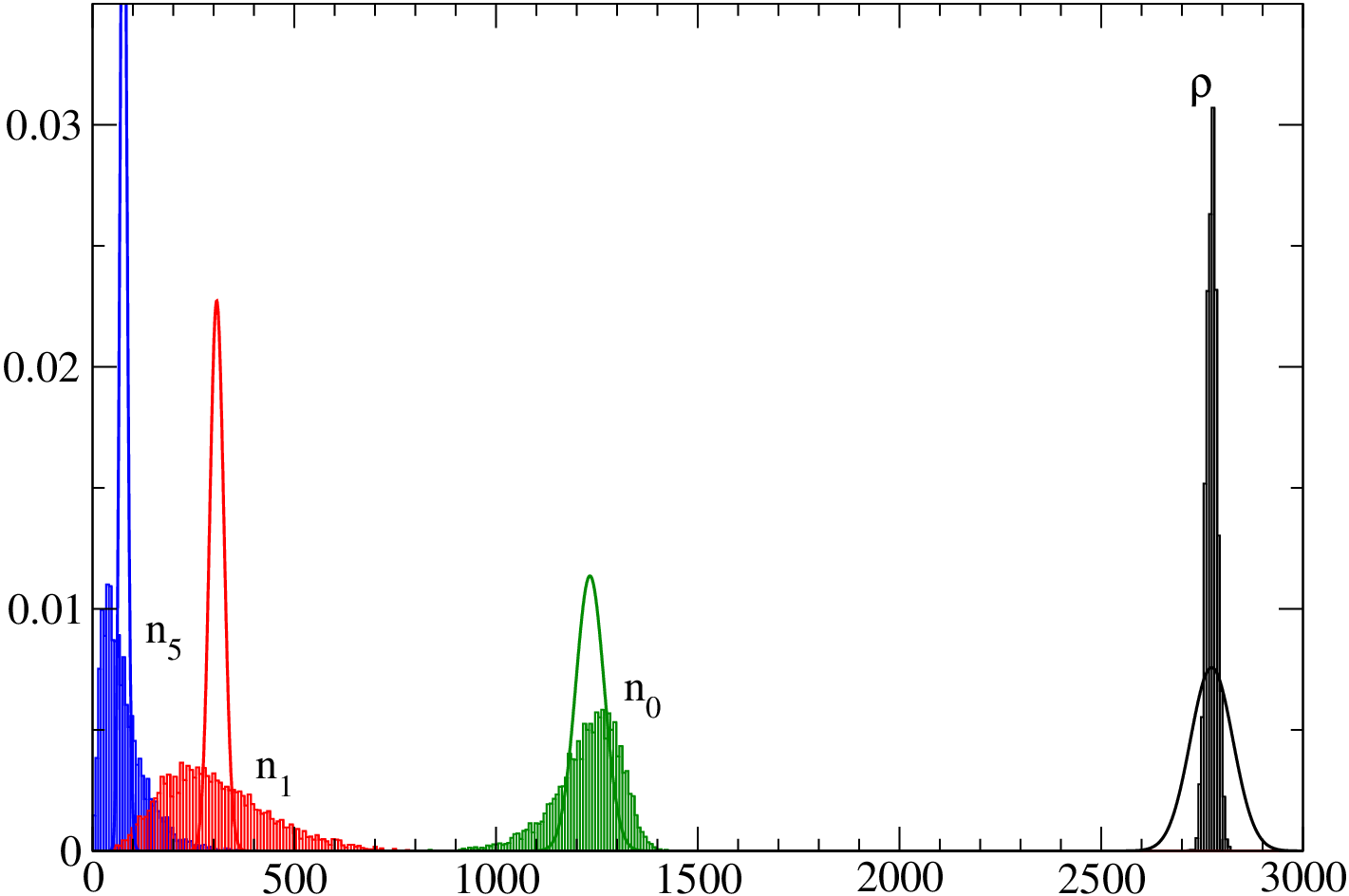}}
  \caption{Distribution of the MDLG occupation numbers $n_i$ and the total density compared to the Poisson distribution for number density $\rho\approx 1.1$ and the lattice spacing is (a) $\Delta x = 12$ and (b) $\Delta x = 50$.}
  \label{fig2}
\end{figure}

For larger densities, where the excluded volume becomes important, the total density will no longer be Poisson distributed, but rather show a narrower distribution. We therefore would also expect the distribution of the lattice gas occupation numbers $n_i$ to similarly deviate from the Poisson distribution. We performed a similar simulation for a denser system of a nominal volume fraction of $\phi=0.49$. The results shown in Figure \ref{fig2}(a) show the narrowing of the $\rho$ distribution, but surprisingly the distribution of rest-particles ($n_0$) appears little changed where as the distributions of particle moving to nearest neighbors ($n_1$) as well as diagonal neighbors ($n_5$) is much wider.

We stated above that we chose $a^2=1/6$, but for each lattice spacing $\Delta x$ there is a time step $\Delta t$ that corresponds to this value. It is therefore essential how the distributions depend on $\Delta x$ (under the assumption $a^2=1/6$), corresponding to a modeling of the system at different scales. Increasing $\Delta x$ will increase the average number of particles per cell. For an ideal system the width of the Poisson distribution will grow only as the square root of the number of particles, making the distribution more peaked for larger number of particles. This means that for larger $\Delta x$ the importance of fluctuations declines. This classical result is found to good approximation for our simulation of a dilute system, as shown in Figure \ref{fig1}(b).

Now for a dense system we show the results for the larger $\Delta x$ in Figure \ref{fig2}(b). The density does indeed show a sharpening of the normalized width, as one would expect. The normalized width of the distributions of the $n_i$, however, does not show the same amount of narrowing. This is a counter intuitive result, which suggests that the importance of fluctuations for the $n_i$ does not diminish as $1/\sqrt{N}$  with increasing $\Delta x$.

To gain an understanding of this phenomenon let us first consider how combining wider distributions for the $n_i$ can lead to a narrower distribution of $\rho$. We can write
\begin{equation}
  \langle(\rho-\rho^{eq})^2\rangle = \sum_i \sum_j \langle(n_i-f_i^{eq})(n_j-f_j^{eq})\rangle.
\end{equation}
For an ideal system the distributions of the $n_i$ are independent. For ideal systems $\langle(n_i-f_i^{eq})(n_j-f_j^{eq})\rangle=f_i^{eq}\delta_{ij}$. But for non-ideal systems the wider distributions for the $n_i$ requires that at least for some $i$ and $j$ these correlations must become quite negative to cancel the widening of the distributions for $i=j$. 

To look at this more quantitatively we can examine the particle displacement probability for different particles $\delta_j = x_j(t)-x_j(t-\Delta t)$. First we can derive the expectation value
\begin{align}
  &\langle n_i\rangle \nonumber\\
  =& \left\langle\sum_{k=1}^N \Delta_x[x_k(t)] \Delta_{x-v_i}[x_k(t-\Delta t)]\right\rangle \nonumber\\
  =&\int dx_1\int d\delta_1 \cdots \int d\delta_N P^N(x_1,\delta_1,\cdots,x_N,\delta_N) \nonumber\\&\sum_{k=1}^N\Delta_x(x_k) \Delta_{x-v_i}(x_k-\delta_k)\nonumber\\
  =&N \int dx_1 \int d\delta_1 \cdots \int d\delta_N P^N(x_1,\delta_1,\cdots,x_N,\delta_n) \nonumber\\&\Delta_x(x_1) \Delta_{x-v_i}(x_1-\delta_1)\nonumber\\
  =&N \int dx_1\int d\delta_1 P^1(x_1,\delta_1) \Delta_x(x_1) \Delta_{x-v_i}(x_1-\delta_1)\nonumber\\
  =& f_i^{eq}
\end{align}
where we introduced the useful (but not very common) N-particle distribution function for the {\em displacements} of particles during a time-step $\Delta t$. 
We see that the expectation value of the distribution is entirely dependent on the one-particle distribution function. This is the reason that Parsa \textit{et al.} \cite{parsa2018validity} found that the equilibrium distribution depends on the non-dimensional mean squared displacement $a^2$ only.

The probability for finding a specific occupation number $n_i$ is then given by
\begin{align}
  & P(n_i) \nonumber\\
  =& \int d x_1 \int d\delta_1\cdots \int d\delta_N P^N(x_1,\delta_1, \cdots,x_N,\delta_N) \nonumber\\&
  \Theta(n_i;\{x_1,\delta_1,\cdots,x_N,\delta_N\} ) \nonumber\\
  =& 
    \int d x_1 \int d\delta_1\cdots \int d\delta_N P^N(x_1,\delta_1, \cdots,x_N,\delta_N) \nonumber\\&
    \left(\begin{array}{c} N\\n_i\end{array}\right)
       \Theta(n_i;\{x_1,\delta_1,\cdots,x_{n_1},\delta_{n_1}\} ) 
\end{align}
where we define a binary flag that is one if we have the occupation number $n_i(x)$ as
\begin{align}
  &\Theta(n_i;\{x_1,\delta_1,\cdots,x_M,\delta_M\})\nonumber\\
  =&\left\{
  \begin{array}{cl}
    1 & \mbox{ if }\sum_{k=1}^M \Delta_x(x_k)\Delta_{x-v_i}(x_k-\delta_k)\equiv n_i\\
    0 & \mbox{ else }
    \end{array}\right.
\end{align}
In the special case of an ideal gas, where the $N$-particle distribution factorizes
\begin{equation}
  P^{N,id}(x_1,\delta_1,\cdots,\delta_N) = \prod_{k=1}^N P^1(x_k,\delta_k)
  \label{factor}
\end{equation}
we have
\begin{align}
  P^{id}(n_i) &= \left(\begin{array}{c} N\\n_i\end{array}\right) \left(\frac{f_i^{eq}}{N}\right)^{n_i} \left(1-\frac{f_i^{eq}}{N}\right)^{N-n_i}\nonumber\\
    &\approx \exp(-f_i^{eq})\frac{(f_i^{eq})^{n_i}}{n_i!}
\end{align}
where the last step is the familiar transition from the binomial distribution to the Poisson distribution in the limit of large $N$.
We can shed light on the expected width of the distribution for the non-ideal case by reducing the expression to one for the two-particle distribution function:
\begin{align}
  & \langle n_i(x) n_j(y) \rangle
  \nonumber\\=&\left\langle \sum_{k=1}^N \Delta_x(x_k)\Delta_{x-v_i}(x_k-\delta_k)\sum_{l=1}^N\Delta_y(x_l)\Delta_{y-v_j}(x_l-\delta_l)\right\rangle
  \nonumber\\=&
  \sum_{k,l}\int d x_1 \int d \delta_1 \cdots\int d\delta_N P^N(x_1,\delta_1,\cdots,x_N,\delta_N)
  \nonumber\\&\Delta_x(x_k)\Delta_{x-v_i}(x_k-\delta_k)\Delta_y(x_l)\Delta_{y-v_j}(x_l-\delta_l)\nonumber\\
  =& (N^2-N) \int d x_1 \int d \delta_1 \int dx_2\int d\delta_2 P^2(x_1,\delta_1,x_2,\delta_2)
  \nonumber\\&\Delta_x(x_1)\Delta_{x-v_i}(x_1-\delta_1)\Delta_y(x_2)\Delta_{y-v_j}(x_2-\delta_2)\nonumber\\&
  + N \int d x_1 \int d \delta_1 P^1(x_1,\delta_1)
  \nonumber\\&\Delta_x(x_1)\Delta_{x-v_i}(x_1-\delta_1)\Delta_y(x_1)\Delta_{y-v_i}(x_1-\delta_1)\nonumber\\
  =& (N^2-N) \int d x_1 \int d \delta_1 \int dx_2\int d\delta_2 P^2(x_1,\delta_1,x_2,\delta_2)
  \nonumber\\&\Delta_x(x_1)\Delta_{x-v_i}(x_1-\delta_1)\Delta_y(x_2)\Delta_{y-v_j}(x_2-\delta_2)\nonumber\\&
  + f_i^{eq} \delta_{ij} \delta_{xy}
  \label{ninjeqn}
\end{align}
where, for an ideal gas, the two particle probability factorizes (\ref{factor}).
With this we obtain for an ideal gas
\begin{align}
  \langle n_i(x) n_j(y)\rangle^{id} = (1-1/N) f_i^{eq} f_j^{eq}+f_i^{eq} \delta_{ij} \delta_{xy}
\end{align}
Which is the result that one would expect for an independently binomial distributed $n_i$. In the limit of a large system with $N\rightarrow \infty$ the $1/N$ can be neglected, and we obtain the result predicted for a Poisson distributed $n_i$.

\begin{figure}
  \centering
  \subfloat[]{\includegraphics[width=0.9 \columnwidth]{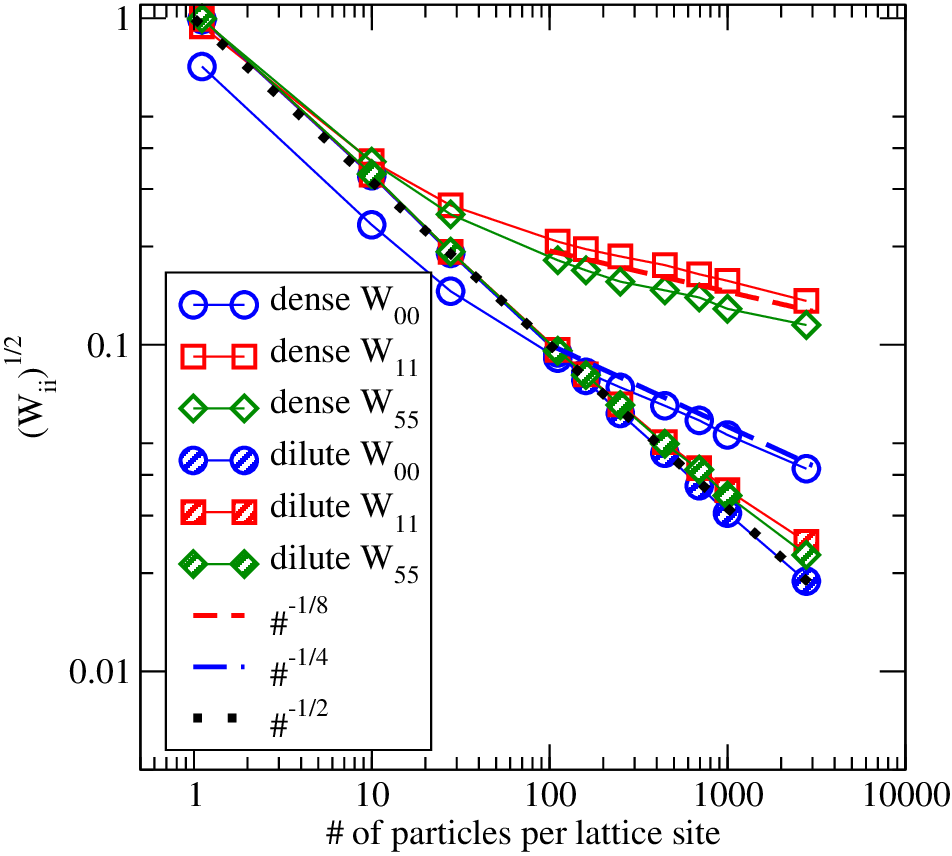}}\\
  \subfloat[]{\includegraphics[width=0.9 \columnwidth]{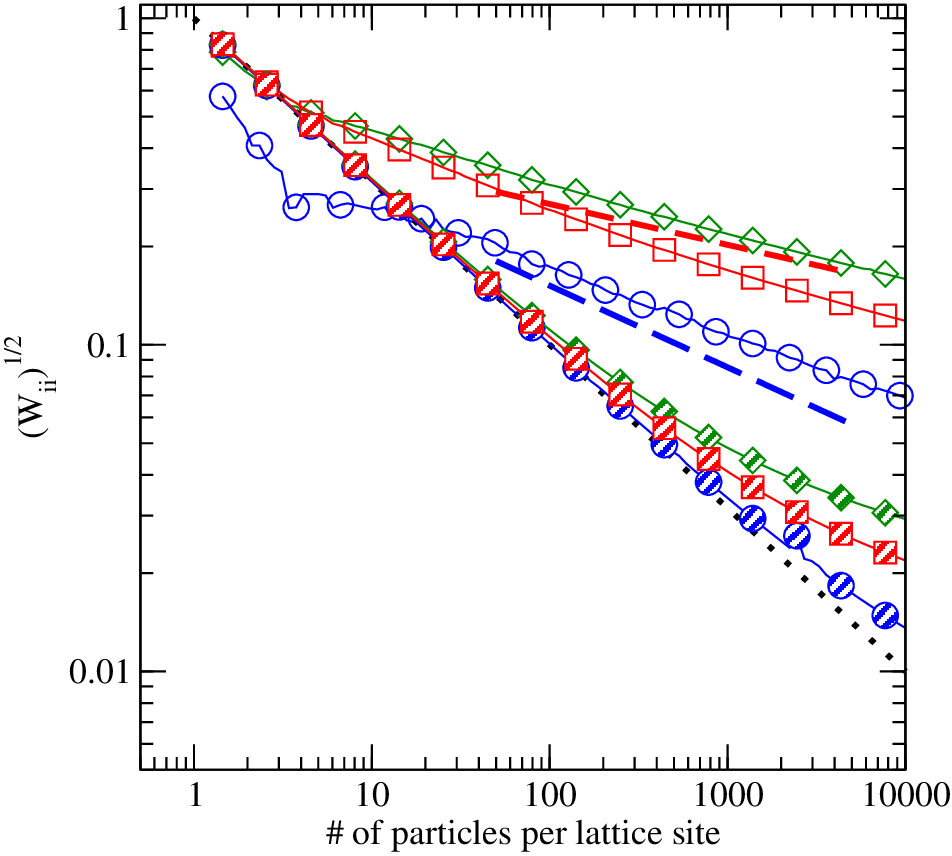}}  
  \caption{Scaled width $\sqrt{W_{ii}}$ of the $n_i$ distributions compared to that of a Poisson distribution for the MDLG system (a) and a theoretical prediction from the simplified two-particle displacement probability of Eq. (\ref{simpP2}) (b). For larger coarse-graining the width of the distributions can be orders of magnitude wider than expected for an ideal gas. This transition is enhanced for denser systems, but appears for all densities. The theoretical result obtained by numerical evaluation with Mathematica shows some numerical wiggles, indicating that the software has some difficulties evaluating the required four dimensional integral.}
    \label{fig5}
\end{figure}

For dilute gases the representation of the particles as ideal particles works quite well. We define the scaled squared width of the actual distribution as
\begin{equation}
  W_{ii}=\frac{\langle (n_i(x)-f_i^{eq})^2\rangle}{\rho^{eq}f_i^{eq}}.
\end{equation}
For a Poisson distribution this squared width will decrease as $1/\rho^{eq}$. 
We plot $\sqrt{W_{ii}}$ in Figure \ref{fig5} for an ideal gas, a real dilute gas and a dense gas as a function of the number of average number particles per lattice site. This is accomplished by increasing the lattice spacing $\Delta x$, and increasing the time step $\Delta t$ at the same time such that $a^2=1/6$ as in Eq. (\ref{a2eqn}).  Compared to the result expected for a Poisson distribution the dilute gas agree quite well up to about a thousand particles per lattice site, but after that we see that the width appears to fail to decrease as fast as expected for an ideal gas. This is in agreement with the results for a dilute gas in Figure \ref{fig1}, which showed that the distributions agreed reasonably well with the prediction for a Poisson distribution.

For the dense gas the scaled width of the distribution diverges from the ideal gas case already at about 10 particles per lattice site, and decays much more slowly for larger numbers of particles. This implies that the importance of fluctuations does no longer decay as $1/\sqrt{\rho^{eq}}$. Therefore increasing the coarse-graining by choosing a larger lattice spacing $\Delta x$ will not diminish as rapidly as would be expected in standard Statistical Mechanics. The difference is striking: instead of decaying as $(\rho^{eq})^{-1/2}$ the rest-particles scale approximately as $(\rho^{eq})^{-1/4}$ and moving particle density only as $(\rho^{eq})^{-1/8}$. This result may seem counter intuitive: basic Statistical Mechanics would seem to demand that if we continue to double the lattice spacing $\Delta x$, the added components should eventually become independent. This argument, however, misses the important point that we are keeping $a^2=1/6$, so that the time step $\Delta t$ also increases.

To understand why the correlations do not vanish more quickly when scaling up the coarse-graining of the system let us consider the probability of the particle displacements. Firstly it is important to realize that instantaneous velocities in equilibrium remain uncorrelated
\begin{equation}
  P^2(x_1,v_1,x_2,v_2) \propto  P(v_1) P(v_2) g(|x_1-x_2|)
  \label{pfact}
\end{equation}
where $g(r)$ is the pair correlation function.
The same, however, is not true for displacement probabilities $P_2(x_k,\delta_k,x_l,\delta_l)$. Since these displacements can be written as
\begin{equation}
  \delta_k = \int_0^{\Delta t} v_k(t) dt
\end{equation}
the probability contains information about velocity cross correlations that are considered in the context of cross-diffusion constants. To understand this better we have to consider $P(x_1,\delta_1,x_2,\delta_2)$ of Eq.(\ref{ninjeqn}). For two-dimensional systems considered in this letter this probability depends on 8 variables. Because of overall translational and rotational invariance this could be reduced to 5 variables. However, this is still too large to be efficiently probed in our simulations. It is reasonable to expect that for particles sufficiently far apart the displacement probability will factorize:
\begin{equation}
  \lim_{|x_2-x_1|\rightarrow \infty} P^2(x_1,\delta_1,x_2,\delta_2) = P^1(x_1,\delta_1)P^1(x_2,\delta_2)
  \label{fact}
\end{equation}
For particles closer together, however, the probability no longer needs to factorize.  It is well known that, particularly for longer times the time-correlator for velocities decays only algebraically \cite{dorfman1994generic}. It is reasonable to expect a similar effect for velocities of particles that have a small spatial separation. A comprehensive numerical evaluation of these correlations is outside the scope of this letter, but we did examine $\langle \delta_1 \delta_2\rangle$ as a function of the initial displacement of the particles $|x_2-x_1|$ \cite{parsa2018lattice}. We found that this correlator does appear to decay exponentially with a correlation length $\xi \approx \langle \delta^2\rangle$ with a prefactor that varies remarkably little with density (from 1 to 1.25 for the two extreme densities studied in this letter).

We now pick a length scale for our lattice $\Delta x$ and a corresponding time scale $\Delta t$ such that $a^2=1/6$. This means that $\xi$ measured in terms of $\Delta x$ is nearly constant. We can also determine the dependence of the amplitude on $\Delta x$ and the overall density to get the approximate relation
\begin{equation}
  \frac{\langle \delta_1 \delta_2\rangle(|x_1-x_2|)}{\langle \delta_1^2\rangle} \approx 2 (\Delta x)^{-1/2} \rho^{1/2} \exp\left(-\frac{|x_1-x_2|}{\xi \Delta x}\right)
  \label{corr}
\end{equation}
where $\xi$ now varies slightly from 1 to 1.25 for our density range. There are two caveats to this approximate result: firstly at short distances the normalized correlation of Eq. (\ref{corr}) has to be less than one. Secondly the total expectation of $\langle \delta_1 \delta_2\rangle =0$ because of momentum conservation. In numerical experiments we find that for large $\Delta x$ there is actually a negative correlation, but the value of this correlation function for large $\Delta x$ is not of interest here. Even with these caveats this result implies that the non-ideal part of the two-particle probability also decays exponentially with correlation length $\xi$. We can now make the following Ansatz that will recover both the factorization of Eq. (\ref{pfact}) and the correlation of Eq. (\ref{corr}):
\begin{align}
    P&(x_1,\delta_1,x_2,\delta_2) \propto g(r)\nonumber\\
& \times\exp\left(-\frac{(\delta_1+\delta_2)^2}{4\sigma_+^2(r)}\right)
  \exp\left(-\frac{(\delta_1-\delta_2)^2}{4\sigma_-^2(r)}\right)
   \label{simpP2}
\end{align}
with
\begin{align}
  r&=|x_1-x_2|\\
  \sigma_+(r)&=a\Delta x\sqrt{1+\frac{\langle \delta_1 \delta_2\rangle (r)}{\langle \delta_1 \delta_1\rangle}}\\
  \sigma_-(r)&=a\Delta x\sqrt{1-\frac{\langle \delta_1 \delta_2\rangle (r)}{\langle \delta_1 \delta_1\rangle}}
\end{align}
This is the simplest two particle distribution function that recovers both Eq. (\ref{fact}) and Eq. (\ref{corr}). We can use numerical integrations to estimate the occupation number correlators in Eq. (\ref{ninjeqn}). The numerical integration was performed using Mathematica, using the rough two particle correlation function
\begin{equation}
  g(r) = \frac{1}{2} \tanh\left(\frac{|x_1-x_2|-r_c}{0.03}\right)+\frac{1}{2}
\end{equation}
and a notebook is included in the supplemental material. The results of this numerical integration is shown in Fig. \ref{fig5}, and it qualitatively recovers the result of the direct simulations.

In conclusion we have found that apparent equal-time fluctuations in coarse grained models contain time-correlations, which can significantly alter the scaling of the fluctuations. It is well known that while equal-time fluctuations in equilibrium systems decay exponentially, different-time fluctuations only decay algebraically \cite{dorfman1994generic} so that the contribution of different-time correlations can be significant. This does not change the short-ranged nature of these fluctuations, as can be seen in Eq. (\ref{corr}) but the effects are nevertheless significant. We find that the magnitude of fluctuations can be orders of magnitude larger for the coarse-grained system variables when compared to quantities that are obtained from equal-time correlators like density or momentum fluctuations. This property of mesoscopic methods has not been fully appreciated to date. A long history of previous lattice gas approaches made a Markov approximation, which implies that occupation numbers were viewed as instantaneous quantities. Such a view is in contrast to our lattice gas which is a true coarse-graining of a reality (as represented by a MD simulation), and has fundamentally different properties.  This interpretation of a lattice gas as a coarse-grained model represents a shift in perception, and we believe that our results will facilitate the development of more realistic coarse-grained fluctuating methods.

Lastly we would like to emphasize that the large fluctuations observed in this letter should not be thought of as restricted to lattice gas approaches, but that these will appear in all coarse-grained approaches, like those mentioned in the introduction.

\bibliography{AW,MDLGFluctLett,IntegerLG,MCLG}

\end{document}